# Anisotropic transport at the LaAlO$_3$/SrTiO$_3$ interface explained by microscopic imaging of channel-flow over SrTiO$_3$ domains

Yiftach Frenkel[1], Noam Haham[1], Yishai Shperber[1], Chris Bell[2], Yanwu Xie[3,4], Zhuoyu Chen[5], Yasuyuki Hikita[3], Harold Y. Hwang[3,5], and Beena Kalisky[1,*]

1. Bar-Ilan University, Department of Physics and Institute of Nanotechnology and Advanced Materials, Ramat-Gan, Israel
2. H. H. Wills Physics Laboratory, University of Bristol, Tyndall Avenue, Bristol, BS8 1TL, UK
3. Stanford Institute for Materials and Energy Sciences, SLAC National Accelerator Laboratory, Menlo Park, California 94025, USA
4. Department of Physics, Zhejiang University, Hangzhou, 310027, China
5. Department of Applied Physics, Geballe Laboratory for Advanced Materials, Stanford University
476 Lomita Mall, Stanford University, Stanford, California, 94305, USA
* beena@biu.ac.il

**Abstract**: Oxide interfaces, including the LaAlO$_3$/SrTiO$_3$ interface, have been a subject of intense interest for over a decade due to their rich physics and potential as low dimensional nanoelectronic systems. The field has reached the stage where efforts are invested in developing devices. It is critical now to understand the functionalities and limitations of such devices. Recent scanning probe measurements of the LaAlO$_3$/SrTiO$_3$ interface have revealed locally enhanced current flow and accumulation of charge along channels related to SrTiO$_3$ structural domains. These observations raised a key question regarding the role these modulations play in the macroscopic properties of devices. Here we show that the microscopic picture, mapped by scanning superconducting quantum interference device, accounts for a substantial part of the macroscopically measured transport anisotropy. We compared local flux data with transport values, measured simultaneously, over various SrTiO$_3$ domain configurations. We show a clear relation between maps of local current density over specific domain configurations and the measured anisotropy for the same device. The domains divert the direction of current flow, resulting in a direction dependent resistance. We also show that the modulation can be significant and that in some cases up to 95% of the current is modulated over the channels. The orientation and distribution of the SrTiO$_3$ structural domains change between different cooldowns of the same device or when electric fields are applied, affecting the device behavior. Our results, highlight the importance of substrate physics, and in particular, the role of structural domains, in controlling electronic properties of LaAlO$_3$/SrTiO$_3$ devices. Further, these results point to new research directions, exploiting the STO domains ability to divert or even carry current.

The observed emergence of new low-dimensional states of matter at transition metal oxide heterointerfaces has motivated intense interest. Considerable efforts have been invested in order to understand both the basic physical phenomena in these systems, and their potential for future devices [1–4]. The {100} interface between LaAlO$_3$ and TiO$_2$ terminated SrTiO$_3$ (LAO/STO), exhibits many fascinating properties such as quasi-two-dimensional electron transport with high electron mobility [5], two-dimensional superconductivity at low temperatures [3,6,7], magnetism [8,9] and superconductivity that co-exist [10–13], and electric field-tuned metal-insulator and superconductor-insulator phase transitions [3,7,14]. Various theoretical and experimental considerations suggest that the electron gas is formed by the occupation of several sub-bands [15–17]. Anisotropic transport in strong magnetic fields has been suggested to arise naturally as a consequence of the orbital character of the quantum sub-bands [18–20].

Concomitant with these electrical transport measurements, scanning probe studies have been used to locally visualize the emergent magnetism in the system[10,11,21] and most notably demonstrate local variations in electronic properties associated with the formation of tetragonal domains in the STO at low temperatures [22,23]. A central question to resolve is the inter-relation between these measurements. An influence of local variations on the overall sample transport can have implications on the potential of the LAO/STO interface for future electronics. In this work, we show that transport in patterned LAO/STO-based devices is highly anisotropic, and controlled by the STO domain configuration. We quantify the relation between the local variations of electronic properties and the transport behavior, and discuss obstacles in realization of devices.

STO is a cubic perovskite at room temperature and undergoes a ferroelastic phase transition at ~105 K [24,25]. In the absence of a symmetry breaking field at the transition the tetragonal STO unit cells can form in any of the three possible crystallographic axes and hence the STO breaks into domains. Measurements taken with a scanning SQUID (Superconducting QUantum Interference Device) showed that current flow at the LAO/STO interface can be enhanced along the <100> and <110> STO cubic crystallographic directions depending on the details of the domain structure [22]. Scanning single-electron transistor data demonstrated modulated electrostatic potential on domain walls, predicting an enhancement of flow along domain walls [23]. Here, we use scanning SQUID to map the microscopic spatial distribution of the current flow and compare it to macroscopic resistance measurements performed simultaneously.

In this study we measured LAO/STO samples with 5 unit cells of LAO patterned as hall bars and square devices (methods). An AC bias current in the range of 10-35 µA and frequency of 223 Hz was driven in the sample during measurement, and its resultant magnetic field was measured by the SQUID (Fig. 1a and methods). We used a lock-in amplifier locked to the external current frequency for noise reduction. We verified that the SQUID response to the applied current was in the linear regime, and independent of frequency in the range of d.c. to ~1.9 kHz. Simultaneously, we measured the resistance using a four-probe technique. Fig. 1b-c show typical magnetic flux measurements taken at different regions of a Hall bar device. We observe magnetic maps which are consistent with a homogeneous current flow (Fig.1b), localized regions of reduced conductivity (Fig.1b) and a stripe-like structure of enhanced conductivity channels (Fig.1c). Fig. 1d shows the magnetic flux as a function of position at specific spatial cuts and the schematic current pattern attributed to them.

Hall bar devices enable precise four probe measurement of conductivity, as the current flows in a defined direction of the device. Our goal is to determine the anisotropy stemming from modulated current flow, similar to fig. 1c. For testing this experimentally, we need to compare between two situations - current flowing in the direction of the stripy modulation and orthogonal to it. This is not possible in a Hall bar configuration (see Fig. 2a-b), as the current flows only in one pre-defined direction. A squared configuration (Fig. 2c-d) is more suitable for this purpose since we can rotate the current direction over the same modulation. The disadvantage of this geometry is the less defined current direction, which may amplify the measured anisotropy compared to a Hall bar configuration. In Fig. 2e we estimate the differences in the anisotropy by numerically solving the continuity current equation with spatially dependent conductance, simulating stripes of different conductivity, width and density. We compare it to the expected anisotropy from an ideal Hall bar with the same conductance pattern. The graph shows the resistance anisotropy for both geometries as a function of the stripes width for 5 stripes and conductivity ratio of 5 (conductivity of stripes/conductivity of bulk). We show that the anisotropy of the square geometry is generally larger than the anisotropy of the Hall bar geometry, with ratio (red) of typically 2.5. We find that for small number or very narrow stripes this difference is much less significant.

Fig. 3c-d shows magnetic maps for two different current directions shown in Fig. 3a-b, respectively. We observe sharp stripe-like features and hole-like features in both cases. Subtraction of a Gaussian low pass filter is shown in Fig. 3e-f; an enhancement of the features is obtained. These features reflect abrupt variations in the local current distribution which are not expected for spatially homogeneous samples as shown in Fig. 3g-h. We extracted the local current distribution from the magnetic field spatial distribution shown in Fig. 3c-d using the methods described in Ref. [26], and the current magnitude distribution is shown in Fig. 3i-j. We observe a clear deviation from the current distribution expected for a homogeneous sample (Fig. 3k). Again we observe wide stripes in which the current is large and thin stripes in which the current is small. We also note hole-like regions of reduced current. The local variation in current reflects local change in the sample conductivity; i.e. the wide stripes are regions with higher conductivity than the thin stripes. Simultaneously, we measured the global resistance of the sample revealing a surprising result; we obtained an extremely strong resistance anisotropy. The resistance measured with the current flowing horizontally in Fig. 3a, nominally perpendicular to the stripes, was $R_\perp = V_\perp/I_\perp$ = 50 Ω, nearly four times larger than at the case in which current was flowing mostly vertically (Fig. 3b), along the stripes, $R_\parallel$ = 14 Ω. The effect of anisotropy is also manifested in the local current distribution; in Fig. 3i the stripes divert the current to flow at longer

paths, compared to the homogeneous medium case (Fig. 3k). In Fig. 3j more of the current takes the shorter paths, resulting in a smaller voltage drop and correspondingly lower resistance.

It has been shown that the striped structure in LAO/STO current distribution is correlated with STO domain configuration in the sample [22]. In our case the stripes are nearly parallel to the y-axis of the device (2.7° from the left edge), meaning that all domains are in the <010> direction of the original cubic phase [22,27]. In order to confirm the interplay between the stripe configuration and the sample resistance, we repeated the same experiment after cycling the temperature above 105 K and back to 4K. At ~105 K STO has a structural transition and different domain configurations are expected to form upon cooling [25]. We repeated the experiment 12 times, the results for 7 are shown in Fig. 4. Fig. 4a shows $R_\perp$ and $R_\parallel$ for the different domain configurations which their magnetic maps are shown in Fig. 4b. We observe that when the stripes are along <100> or <010> a strong anisotropy is obtained with the resistance of the current flowing along the stripes significantly lower than the resistance of the current flowing perpendicular to them (Configurations 1,6,7). When the stripes are mostly along <110> or <1-10> directions (Configuration 2,3,4,5) we measure almost no difference between the resistance of the two directions since both $I_\perp$ and $I_\parallel$ nominally flow 45 degrees relative to the stripes. This is also shown in Fig. 4c which shows the anisotropy parameter, defined by $2|R_\perp - R_\parallel|/(R_\perp + R_\parallel)$ for the different domain configurations.

To fully address the connection between the local variations of conductivity and the measured global anisotropy, we focus on the most pronounced case of resistance anisotropy, shown in Fig.3. Converting the local current distribution to a local conductivity map is a rather challenging problem, however, in the case the current flows mostly along the stripes (Fig. 3j) we do not expect a significant variation in the electric field from the homogeneous conductivity case as the stripes effectively act as resistors connected in parallel. Thus, for a rough estimation, we divide the local current distribution (Fig. 3j) by the electric field simulated for a homogeneous medium sample to extract the local conductivity.

The local conductivity map is shown in Fig.5a. We observe stripes of low and high conductivity consistent with the local current distribution map. Using the extracted conductivity map we numerically calculate the resistance for the horizontal and vertical directions. We obtain a ratio between $R_-$ and $R_|$ of 3.8, consistent with the measured ratio of 3.57 in this case. Thus, we determine quantitatively that the major source of the observed anisotropy are the stripes of conductivity modulation, correlated with the STO domains.

We quantified the modulation of current flowing along the domains, for individual stripes, by defining it as the difference between the total current (integrated over the stripe width) and "homogeneous" portion (interpolated between the minima in the current cross section), divided by the total current, as shown in the inset to Fig. 5b. A histogram of current modulation on individual stripes is plotted in Fig. 5b, showing that most of these current modulations are small. However, they can also reach 95% meaning that almost the entire current is diverted by the domains rather than flowing at the expected homogeneous pattern. One of the remaining questions is whether the thin stripes originate from thinner domains with lower conductivity or resolution limited domain walls. For example, in Fig. 3i (dashed circle) higher current flows in the wide stripes, and thus we attribute them to domains, while the thinner stripes with lower current could be attributed to domain walls or to thinner domains with reduced conductivity. In this example we observed reduction of the current in the resolution limited stripes and in other configurations we observe enhancement, thus we cannot determine between the two scenarios. Modulated conduction on domain walls supports recent reports of local piezoelectric response in STO, attributed to domain wall polarity[28]. Polar domain walls could affect conduction by causing accumulation of interface screening charges that counteract depolarization fields [29–31]. The domain wall polarity arises at ~80 K and peaks at 40 K [28], consistent with our observation of modulated current flow in stripes only below ~40 K [22].

The effect of the domain configuration should become more pronounced as devices decrease in size, approaching the characteristic length scale of the channels. Large, millimeter-sized samples are more likely to harbor complicated domain configurations within which the inhomogeneous current flow will statistically average to zero, and therefore we do not expect to detect any significant anisotropy in such samples (note the low anisotropy in Fig. 4 configurations 2 & 3). We expect to observe strong anisotropy in devices smaller than ~100 μm, where the possibility of having domains in only one direction is higher. The effect is also relevant for experiments combining electrostatic gate, where domain configuration is changed due to gate induced domain wall motion[22,23]. Although <100> STO domains move away from the center of an unpatterned samples [23], <110> domains show the opposite behavior[27], and as a result modulated flow over domains is difficult to fully exclude from the sample by gate sequence. In order to eliminate the undesired effects of the domains on transport some measures can be taken, such as cooling the devices in the presence of directional strain or electric fields in order to force the STO to mono-domain formation. For developing an efficient method, deeper understanding of the nature of the movement and nucleation of the domains is needed. Controlling domain configuration and dynamics may also guide applications exploiting the functionalities of the domains, as nanoscale mobile devices[32,33].

In conclusion, we observed strong anisotropy in the resistance of LAO/STO devices, which dramatically changed across thermal cycles in the same device. By microscopically mapping the distribution of the current flow, we correlate this anisotropy with the configuration of structural domains. The strong anisotropy implies that transport values can vary significantly over different cooldowns (different domain configurations) of a certain device. The fact that current is strongly diverted by the domains is crucial for understanding the behavior of smaller devices, and for experiments or functionalities where electrostatic gates are applied. Our results emphasize the role of STO physics in determining interface conductivity, among other factors that control variations in transport properties [34–38], which provides an opportunity to use the diverse physical phenomena of STO to control electronic properties of STO-based heterostructures.

**Methods:**

LAO/STO samples with 5 unit cells of LAO film grown on $TiO_2$-terminated {100} STO substrate were prepared as described in Ref.[3]. Patterned Hall bars and square devices were made by pre-patterning the substrate with ~100 nm thick amorphous $AlO_x$ mask which was lift-off deposited at room temperature with oxygen background pressure of 0.1 mbar using pulsed laser deposition.

The samples were measured in a custom-built piezoelectric-based scanning SQUID microscope with a 1.8 μm diameter pick-up loop [39,40]. We used the scanning SQUID microscope to image magnetic fields from the sample as a function of position. The measured flux is given by $\phi_s = \int g(x,y) B \cdot da$ where the integral is taken over the plane of the SQUID, $g(x,y)$ is the point spread function of the pickup loop, B is the magnetic field produced by the sample and $da$ is the infinitesimal area vector element pointing normal to the plane of the SQUID. The AC magnetism measurements were taken by applying an AC current to the sample and collecting the flux created by currents in the sample using lock in techniques. Each flux image is a convolution of the z component of the magnetic field and the SQUID point spread function. A current carrying wire will appear in our images as a black stripe next to a white stripe. In order to obtain a more intuitive image of the current in the sample we inverted the magnetic field image to current image using methods described in Roth *et al.* [41].

**Acknowledgements**

Y.F., N.H, Y.S and B.K. were supported by the European Research Council grant ERC-2014-STG-639792, Marie Curie Career Integration Grant FP7-PEOPLE-2012-CIG-333799, and Israel Science

Foundation grant ISF-1102/13. Y.X, Z.C, Y.H and H.Y.H were supported by the Department of Energy, Office of Basic Energy Sciences, Division of Materials Sciences and Engineering, and Laboratory Directed Research and Development funding, under contract DE-AC02–76SF00515

**Notes**

The authors declare no competing financial interest.

References
(1)     Zubko, P.; Gariglio, S.; Gabay, M.; Ghosez, P.; Triscone, J.-M. Interface Physics in Complex Oxide Heterostructures. *Annual Review of Condensed Matter Physics*. 2011, pp 141–165.

(2)     Mannhart, J.; Schlom, D. G. Oxide Interfaces--an Opportunity for Electronics. *Science* **2010**, *327* (5973), 1607–1611.

(3)     Bell, C.; Harashima, S.; Kozuka, Y.; Kim, M.; Kim, B. G.; Hikita, Y.; Hwang, H. Y. Dominant Mobility Modulation by the Electric Field Effect at the LAO/STO Interface. *Phys. Rev. Lett.* **2009**, *103* (22), 1–4.

(4)     Hwang, H. Y.; Iwasa, Y.; Kawasaki, M.; Keimer, B.; Nagaosa, N.; Tokura, Y. Emergent Phenomena at Oxide Interfaces. *Nat Mater* **2012**, *11* (2), 103–113.

(5)     Ohtomo, A.; Hwang, H. Y. A High-Mobility Electron Gas at the LaAlO3/SrTiO3 Heterointerface. *Nature* **2004**, *427* (6973), 423–426.

(6)     Reyren, N.; Thiel, S.; Caviglia, A. D.; Kourkoutis, L. F.; Hammerl, G.; Richter, C.; Schneider, C. W.; Kopp, T.; Rüetschi, A.-S.; Jaccard, D.; Gabay, M.; Muller, D. A.; Triscone, J.-M.; Mannhart, J. Superconducting Interfaces between Insulating Oxides. *Science* **2007**, *317* (5842), 1196–1199.

(7)     Caviglia, A. D.; Gariglio, S.; Reyren, N.; Jaccard, D.; Schneider, T.; Gabay, M.; Thiel, S.; Hammerl, G.; Mannhart, J.; Triscone, J.-M. Electric Field Control of the LaAlO3/SrTiO3 Interface Ground State. *Nature* **2008**, *456* (7222), 624–627.

(8)     Lee, J.-S.; Xie, Y. W.; Sato, H. K.; Bell, C.; Hikita, Y.; Hwang, H. Y.; Kao, C.-C. Titanium Dxy Ferromagnetism at the LAO/STO Interface. *Nat Mater* **2013**, *12* (8), 703–706.

(9)     Brinkman, A.; Huijben, M.; van Zalk, M.; Huijben, J.; Zeitler, U.; Maan, J. C.; van der Wiel, W. G.; Rijnders, G.; Blank, D. H. A.; Hilgenkamp, H. Magnetic Effects at the Interface between Non-Magnetic Oxides. *Nat. Mater.* **2007**, *6* (7), 493–496.

(10)    Kalisky, B.; Bert, J. A.; Klopfer, B. B.; Bell, C.; Sato, H. K.; Hosoda, M.; Hikita, Y.; Hwang, H. Y.; Moler, K. A. Critical Thickness for Ferromagnetism in LAO/STO Heterostructures. *Nat. Commun.* **2012**, *3* (May), 922.

(11)    Bert, J. A.; Kalisky, B.; Bell, C.; Kim, M.; Hikita, Y.; Hwang, H. Y.; Moler, K. A. Direct Imaging of the Coexistence of Ferromagnetism and Superconductivity at the LAO/STO Interface. *Nat. Phys* **2011**, *7* (September), 2079.

(12)    Ariando; Wang, X.; Baskaran, G.; Liu, Z. Q.; Huijben, J.; Yi, J. B.; Annadi, A.; Barman, A. R.; Rusydi, A.; Dhar, S.; Feng, Y. P.; Ding, J.; Hilgenkamp, H.; Venkatesan, T. Electronic Phase Separation at the LaAlO3/SrTiO3 Interface. *Nat Commun* **2011**, *2*, 188.

(13) Dikin, D. A.; Mehta, M.; Bark, C. W.; Folkman, C. M.; Eom, C. B.; Chandrasekhar, V. Coexistence of Superconductivity and Ferromagnetism in Two Dimensions. *Phys. Rev. Lett.* **2011**, *107* (July), 056802.

(14) Thiel, S.; Hammerl, G.; Schmehl, A.; Schneider, C. W.; Mannhart, J. Tunable Quasi-Two-Dimensional Electron Gases in Oxide Heterostructures. *Science* **2006**, *313* (5795), 1942–1945.

(15) Ben Shalom, M.; Ron, A.; Palevski, A.; Dagan, Y. Shubnikov-de Haas Oscillations in SrTiO3/LaAlO3 Interface. *Phys. Rev. Lett.* **2010**, *105*, 206401.

(16) Joshua, A.; Pecker, S.; Ruhman, J.; Altman, E.; Ilani, S. A Universal Critical Density Underlying the Physics of Electrons at the LAO/STO Interface. *Nat Commun* **2012**, *3*, 1129.

(17) Xie, Y.; Bell, C.; Kim, M.; Inoue, H.; Hikita, Y.; Hwang, H. Y. Quantum Longitudinal and Hall Transport at the LAO/STO Interface at Low Electron Densities. *Solid State Commun.* **2014**, *197*, 25–29.

(18) Ben Shalom, M.; Tai, C. W.; Lereah, Y.; Sachs, M.; Levy, E.; Rakhmilevitch, D.; Palevski, A.; Dagan, Y. Anisotropic Magnetotransport at the LAO/STO Interface. *Phys. Rev. B* **2009**, *80*, 140403.

(19) Annadi, A.; Huang, Z.; Gopinadhan, K.; Wang, X. R.; Srivastava, A.; Liu, Z. Q.; Ma, H. H.; Sarkar, T. P.; Venkatesan, T.; Ariando. Fourfold Oscillation in Anisotropic Magnetoresistance and Planar Hall Effect at the LAO/STO Heterointerfaces: Effect of Carrier Confinement and Electric Field on Magnetic Interactions. *Phys. Rev. B* **2013**, *87*, 220405.

(20) Caviglia, A. D.; Gabay, M.; Gariglio, S.; Reyren, N.; Cancellieri, C.; Triscone, J. M. Tunable Rashba Spin-Orbit Interaction at Oxide Interfaces. *Phys. Rev. Lett.* **2010**, *104*, 126803.

(21) Bert, J. A.; Nowack, K. C.; Kalisky, B.; Noad, H.; Kirtley, J. R.; Bell, C.; Sato, H. K.; Hosoda, M.; Hikita, Y.; Hwang, H. Y.; Moler, K. A. Gate-Tuned Superfluid Density at the Superconducting LAO/STO Interface. *Phys. Rev. B* **2012**, *86*, 060503.

(22) Kalisky, B.; Spanton, E. M.; Noad, H.; Kirtley, J. R.; Nowack, K. C.; Bell, C.; Sato, H. K.; Hosoda, M.; Xie, Y.; Hikita, Y.; Woltmann, C.; Pfanzelt, G.; Jany, R.; Richter, C.; Hwang, H. Y.; Mannhart, J.; Moler, K. A. Locally Enhanced Conductivity due to the Tetragonal Domain Structure in LAO/STO Heterointerfaces. *Nat. Mater.* **2013**, *12* (10), 1091.

(23) Honig, M.; Sulpizio, J. A.; Drori, J.; Joshua, A.; Zeldov, E.; Ilani, S. Local Electrostatic Imaging of Striped Domain Order in LAO/STO. *Nat. Mater.* **2013**, *12* (12), 1112–1118.

(24) Cowley, R. A. The Phase Transition of Strontium Titanate. *Phil. Trans. R. Soc. Lond. A* **1996**, *354*, 2799–2814.

(25) Cowley, R. A. Lattice Dynamics and Phase Transitions of Strontium Titanate. *Phys. Rev.* **1964**, *134* (4A).

(26) Roth, B. J.; Sepulveda, N. G.; Wikswo, J. P. Using a Magnetometer to Image a Two-dimensional Current Distribution. *J. Appl. Phys.* **1989**, *65* (1), 361.

(27) Erlich, Z.; Frenkel, Y.; Drori, J.; Shperber, Y.; Bell, C.; Sato, H. K.; Hosoda, M.; Xie, Y.; Hikita, Y.; Hwang, H. Y.; Kalisky, B. Optical Study of Tetragonal Domains in LaAlO3/SrTiO3. *J. Supercond. Nov. Magn.* **2015**, *28* (3), 1017–1020.

(28) Salje, E. K. H.; Aktas, O.; Carpenter, M. A.; Laguta, V. V.; Scott, J. F. Domains within Domains and Walls within Walls: Evidence for Polar Domains in Cryogenic SrTiO3. *Phys. Rev. Lett.* **2013**, *111* (24), 247603.

(29) Kumar, A.; Kalinin, S. V; Sokolov, A.; Tsymbal, E. Y.; Rzchowski, M. S.; Gruverman, A.; Eom, C. B. Switchable Induced Polarization in LaAlO 3 /SrTiO 3 Heterostructures. **2012**.

(30) Niranjan, M. K.; Wang, Y.; Jaswal, S. S.; Tsymbal, E. Y. Prediction of a Switchable Two-Dimensional Electron Gas at Ferroelectric Oxide Interfaces. *Phys. Rev. Lett.* **2009**, *103* (1), 1–4.

(31) Wang, Y.; Niranjan, M. K.; Jaswal, S. S.; Tsymbal, E. Y. First-Principles Studies of a Two-Dimensional Electron Gas at the Interface in Ferroelectric Oxide Heterostructures. *Phys. Rev. B - Condens. Matter Mater. Phys.* **2009**, *80* (16), 1–10.

(32) Catalan, G.; Seidel, J.; Ramesh, R.; Scott, J. F. Domain Wall Nanoelectronics. *Rev. Mod. Phys.* **2012**, *84* (1), 119–156.

(33) Salje, E. K. H. Domain Boundary Engineering – Recent Progress and Many Open Questions. *Phase Transitions* **2013**, *86* (1), 2–14.

(34) Thiel, S.; Schneider, C. W.; Kourkoutis, L. F.; Muller, D. a.; Reyren, N.; Caviglia, A. D.; Gariglio, S.; Triscone, J. M.; Mannhart, J. Electron Scattering at Dislocations in LaAlO3/SrTiO3 Interfaces. *Phys. Rev. Lett.* **2009**, *102* (4), 166804.

(35) Siemons, W.; Koster, G.; Yamamoto, H.; Geballe, T. H.; Blank, D. H. a; Beasley, M. R. Experimental Investigation of Electronic Properties of Buried Heterointerfaces of LAO on STO. *Phys. Rev. B* **2007**, *76* (15), 155111.

(36) Seri, S.; Shimshoni, E.; Paetel, S.; Mannhart, J.; Klein, L. Angular Dependence of the Magnetoresistance of the LAO/STO Interface. *IEEE Trans. Magn.* **2010**, *46* (6), 1630–1632.

(37) Seri, S.; Schultz, M.; Klein, L. Interplay between Sheet Resistance Increase and Magnetotransport Properties in LAO/STO. *Phys. Rev. B* **2012**, *86* (8), 085118.

(38) Seri, S.; Schultz, M.; Klein, L. Thermally Activated Recovery of Electrical Conductivity in LAO/STO. *Phys. Rev. B* **2013**, *87* (12), 125110.

(39) Huber, M. E.; Koshnick, N. C.; Bluhm, H.; Archuleta, L. J.; Azua, T.; Björnsson, P. G.; Gardner, B. W.; Halloran, S. T.; Lucero, E. A.; Moler, K. A. Gradiometric Micro-SQUID Susceptometer for Scanning Measurements of Mesoscopic Samples. *Rev. Sci. Instrum.* **2008**, *79* (5), 53704.

(40) Gardner, B. W.; Wynn, J. C.; Björnsson, P. G.; Straver, E. W. J.; Moler, K. A.; Kirtley, J. R.; Ketchen, M. B. Scanning Superconducting Quantum Interference Device Susceptometry. *Rev. Sci. Instrum.* **2001**, *72* (5), 2361.

(41) Roth, B. J.; Sepulveda, N. G.; Wikswo, J. P. Using a Magnetometer to Image a Two-Dimensional Current Distribution. *J. Appl. Phys.* **1989**, *65* (1989), 361–372.

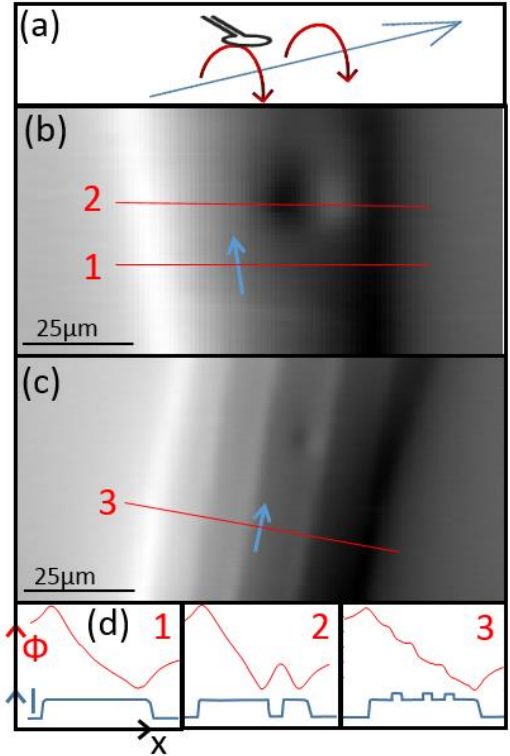

**FIG. 1. Mapping current flow with scanning SQUID microscopy.** (a) The SQUID's pick-up loop is rastered over the sample and captures the z component of the magnetic flux lines generated by the current flow. (b-c) Examples of SQUID images of the magnetic flux generated by in-plane current flow. Blue arrow showing direction of current flow. (d) The cross-sections show the flux responses (red) for the different current distributions (sketched in blue): homogeneous current flow (1), a localized region of reduced conductivity (2), and channels of enhanced current density (3).

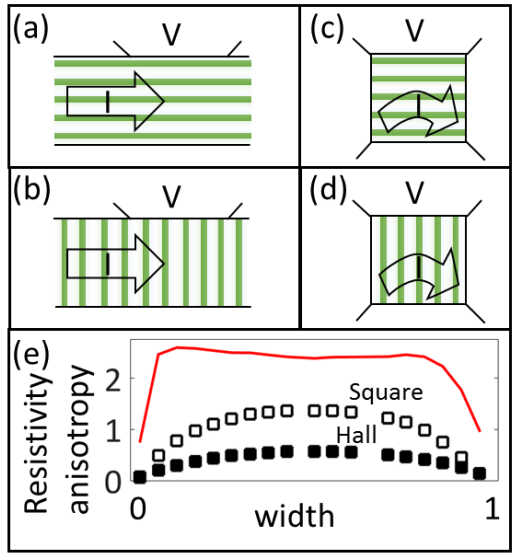

**FIG. 2. Device geometry for studying the anisotropy.** (a-b) Current flowing in a Hall bar geometry with underlying modulations in resistivity. (c-d) Current flowing in a square geometry has a less defined direction, but this geometry is useful for sampling a certain underlying resistivity map from different directions. (e) Calculated resistance anisotropy for a square and an ideal Hall bar geometries as a function of normalized stripes width (width*length*number of stripes/area measured), for conductivity ratio of 5, and 5 stripes. We define the anisotropy as the normalized difference between the resistances on a and b (or c and d for square geometry).

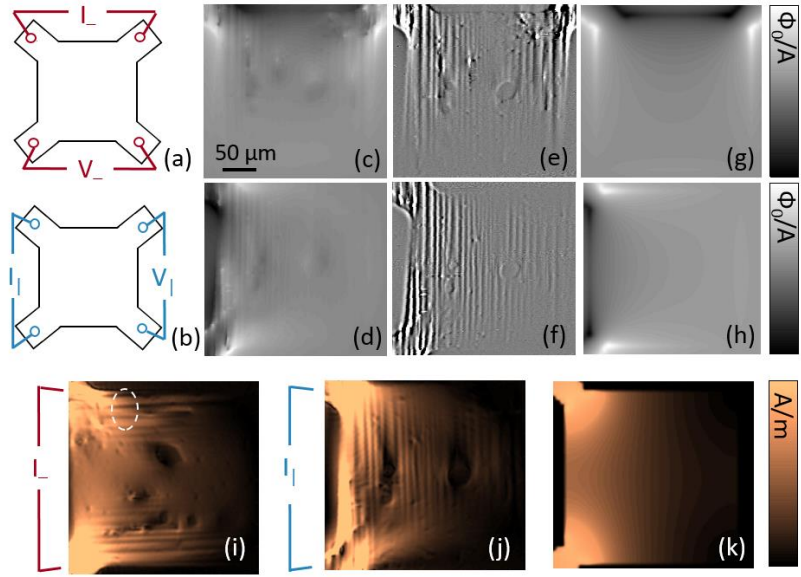

**FIG. 3. SQUID mapping of current flow in different orientations.** (a) Horizontal measurement configuration. (b) Vertical measurement configuration. **(c-f)** Scanning SQUID images of magnetic flux generated by current flow: Raw data (c,d), and after subtracting a low pass Gaussian filter (e,f). (g,h) Simulated magnetic flux for a homogeneous sample by solving the current continuity equation, $\nabla\cdot\mathbf{J}=0$. (i-k) Local distribution of current, extracted from the flux data (i,j), and simulated for a homogeneous sample (k). Linear colorbars span [–5.8 , 5.43] $\Phi_0/A$ (c,d,g,h), [–2.39, 2.43] $\Phi_0/A$ (e,f), [0,1.59] A/m (i-k).

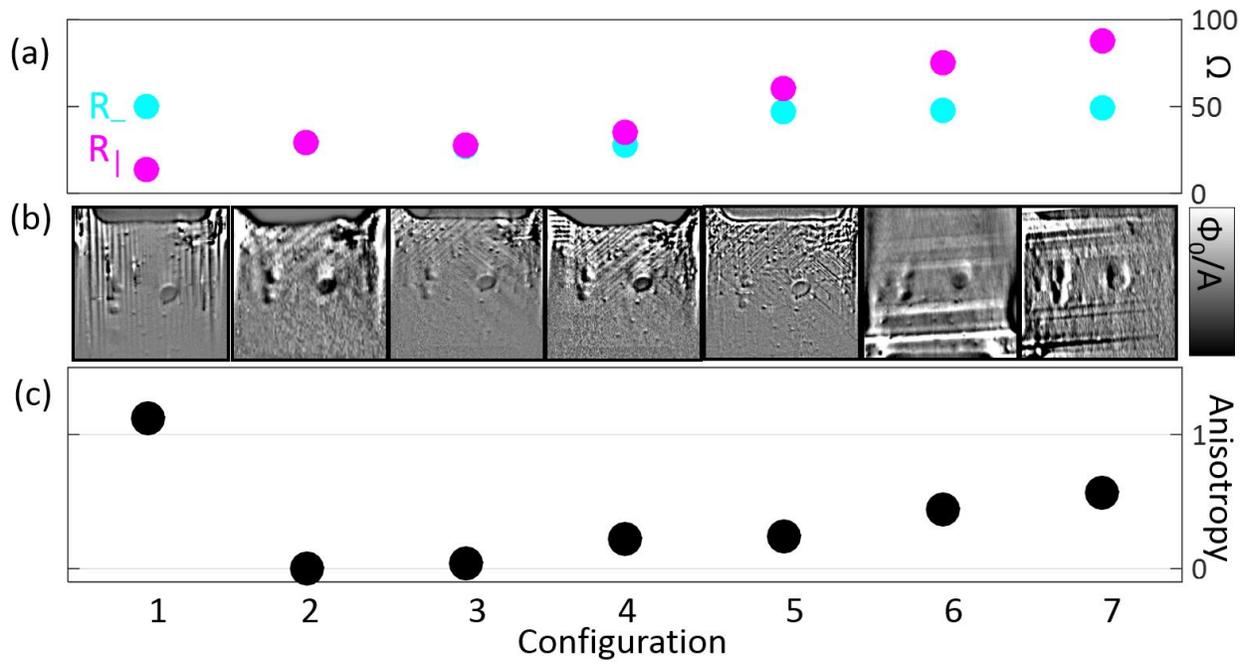

FIG. 4. **Anisotropy of a device as a function of domain configuration**. (a) Resistance measured in the vertical (magenta) and horizontal (cyan) orientations. (b) Flux data for each domain configuration, plotted for one current orientation. (c) Resistance anisotropy, $2|R_- - R_||/(R_- + R_|)$

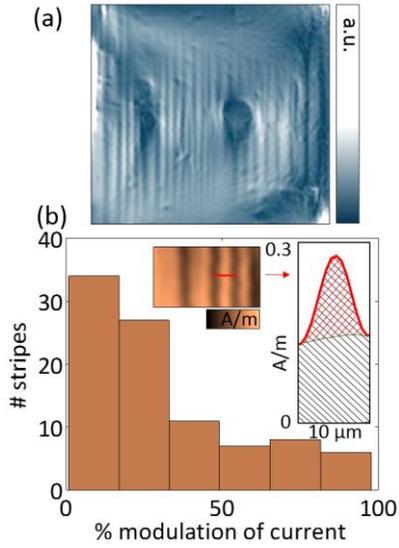

**Fig. 5.** (a) Approximated local conductivity map, extracted from the local current distribution for configuration 1 (configuration 1 is also shown in Fig 2) (b) Histogram of current modulation per stripe, taken from configurations 1-7. (Inset) Part of configuration 1's current distribution map. Red cross-section is taken over one current modulation and plotted to the right. Red area, the current modulation, excluding the estimated homogeneous part of the current. Black area, the entire current in the direction of the domains. The modulation was calculated as the ratio between the areas, with error of 10%.